\documentclass[12pt]{iopart}

\newcommand{\ds}{\displaystyle}
\newcommand{\ov}{\overline}
\def\lesssim{{\
\lower-1.2pt\vbox{\hbox{\rlap{$<$}\lower5pt\vbox{\hbox{$\sim$}}}}\ }} 
\def\gtrsim{{\
\lower-1.2pt\vbox{\hbox{\rlap{$>$}\lower5pt\vbox{\hbox{$\sim$}}}}\ }}
\begin{document}

\title{Towards the robustness of Affleck-Dine mechanism}

\author{Shinta Kasuya}

\address{Department of Information Science, Kanagawa University, \\
Kanagawa 259-1293, Japan}
\ead{kasuya@kanagawa-u.ac.jp}

\begin{abstract}
In this talk we considered the Affleck-Dine mechanism with various types of the K\"ahler 
potential, and investigate whether or not the Affleck-Dine field could acquire a large VEV as 
an initial condition for successful baryogenesis. In addition to a negative Hubble-induced 
mass term, we found examples that large enough Hubble-induced A-terms could also 
develop the minima at large amplitude of the field. It is concluded, therefore, that the 
Affleck-Dine mechanism works for broader classes of the theories. This talk is based 
on Ref.~\cite{KK}. Here I extend the discussion from more generic standpoint, and find 
that essentially there are three cases that large enough A terms develop minima at large 
field values.
\end{abstract}


\section{Introduction}
In the context of supersymmetry (SUSY), a promising candidate of the baryogenesis is 
the Affleck-Dine mechanism \cite{AD,DRT}. It utilizes a scalar field carrying the baryon 
charge, which is called the Affleck-Dine field $\phi$. In particular in the minimal supersymmetric 
standard model, there are a lot of flat directions whose potential vanishes along those 
directions. Since the flat directions consist of squarks and/or sleptons, it is thus  
natural to regard them as the Affleck-Dine field.

During the inflationary stage, the Affleck-Dine field has a large vacuum expectation value (VEV). 
Well after inflation ends, it begins rotation in its potential when the Hubble parameter 
becomes the mass scale of the field, $H \sim m_\phi$. Since the baryon number 
(N\oe ther charge) is given by
\begin{equation}
Q = \int d^3 x \frac{1}{i}\left( \phi \dot{\phi}^* - \dot{\phi} \phi^* \right)
= \frac{1}{2} \int d^3 x \varphi^2 \dot{\theta},
\end{equation}
where $\phi=\varphi e^{i\theta}/\sqrt{2}$, the helical motion implies baryon number production.
In most cases, the Affleck-Dine field feels spatial instabilities, and deforms into Q balls 
\cite{qball1} - \cite{qball7}.
From the decay or evaporation of the formed Q balls, quarks are produced afterwards, 
and we have a baryon asymmetry of the universe in usual sense.

The key ingredient for successful Affleck-Dine baryogensis is how to obtain a large VEV in 
the first place. During inflation, there appears a mass term due to SUSY breaking by the
finite energy density of the inflaton, which is called a Hubble-induced mass term. In supergravity
with the minimal K\"ahler potential, only a positive Hubble-induced mass term arises, which
does not make the field having a large VEV. Therefore, it is usually necessary to have 
nonrenormalizable terms in the K\"ahler potential to obtain a negative Hubble-induced mass 
term, $c_HH^2|\phi|^2$ with $c_H<0$.

In this talk, we show the cases when the field acquires a large VEV due to the negative 
Hubble-induced mass terms for some types of nonminimal K\"ahler potential. On the other hand,
we also consider the opposite situation that the Hubble-induced mass term is positive. Usually 
in this case, the Affleck-Dine field settles down to the origin of the potential, and cannot have a 
large VEV, which implies that the Affleck-Dine mechanism does not work.
The crucial observation, however, reveals that the potential will develop a (local or global) 
minimum at a large amplitude of the field due to Hubble-induced A-terms during and after 
inflation.

\section{Affleck-Dine mechanism due to a negative Hubble-induced mass term}
The potential of the flat direction vanishes only in the SUSY exact limit, and lifted by
SUSY breaking effects and nonrenormalizable operators. The general form of the potential
reads as
\begin{eqnarray}
\label{pot}
V(\phi) &  =  & m_\phi^2 |\phi|^2 + \left( A\frac{\phi^p}{pM_P^{p-3}} + h.c. \right) \nonumber \\
& +&  c_H H^2 |\phi|^2 + \left( a_H H \frac{\phi^q}{qM_P^{q-3}} + h.c. \right) 
+ \lambda^2 \frac{|\phi|^{2(n-1)}}{M_P^{2(n-3)}}.
\end{eqnarray}
The first line represents the effect of (usual) SUSY breaking, while there are Hubble-induced 
mass and A terms in the second line due to finite energy density of the inflaton. 
Here $m_\phi \sim O$(TeV), $A \sim O(m_{3/2})$, $c_H \sim O(1)$, and $a_H \sim O(1)$. 
The last line comes from the nonrenormalizable superpotential of the form, 
$W(\phi)=\lambda \phi^n/nM_P^{n-3}$. In general, $p$, $q$, and $n$ could be different, but
$p=q=n$ in most cases, so we only treat this case hereafter otherwise mentioned.

Since, in order for the Affleck-Dine field to have a large VEV during inflation, the Hubble parameter is necessarily larger than $m_\phi$, the Hubble-induced terms dominate over the 
terms due to (hidden sector) SUSY breaking at that epoch. Thus, the first line of Eq.(\ref{pot}) 
is safely neglected when we consider the dynamics of the flat direction during inflation.

Let us first briefly remind the reader of the usual scenario of the Affleck-Dine mechanism. 
During inflation, the flat direction settles down in the minimum of the potential, which is
determined by the balance of the nonrenormalizable term (the third line of Eq.(\ref{pot})) and
the {\it negetive} Hubble-induced mass term (the first term in the second line of Eq.(\ref{pot})
with $c_H < 0$). Therefore, the amplitude of the minimum is estimated as 
$\varphi_{\rm min} \sim (HM_P^{n-3})^{1/(n-2)}$ where $c_H$, $\lambda \sim O(1)$ are assumed. 
After inflation when the Hubble parameter decreases as large as the mass of the flat direction, 
$H \sim m_\phi$, this minimum disappears and the flat direction begins moving towards the 
origin, the only (global) minimum. At the same time, the Hubble-induced and (hidden sector) 
SUSY breaking A terms become comparable. Since the Hubble parameter becomes also as 
large as the mass scale of the phase direction, 
$m_\theta \sim (A\varphi_{\rm min}^{n-2}/M_P^{n-3})^{1/2}$, the field feels torque due to the 
difference of the minima in the phase direction, and begins helical motion in the potential. 
This is the (dynamical) origin of the CP violation, one of the Sakharov's three conditions for 
baryogenesis. Thus, at the onset of oscillation in the potential, the baryon number density is 
estimated as, 
\begin{equation}
\label{baryon}
n_B \sim \frac{A}{m_\phi} \frac{\varphi_{\rm min}^n}{M_P^{n-3}} 
\sim  \left(\frac{m_\phi}{M_P}\right)^{\frac{n}{n-2}} M_P^3, 
\end{equation}
for $O(1)$ difference of the potential minima in the phase direction due to the usual and 
Hubble-induced A terms, and $A \sim m_\phi$ is used. In this scenario, the key is having a 
negative Hubble-induced mass term in order for the field to acquire a large VEV during 
and after inflation before the Affleck-Dine field starts its rotation.

\section{Hubble-induced mass terms}
In the supergravity the scalar potential is written in terms of superpotential, 
$W$, and K\"ahler potential, $K$, as
\begin{equation}
V=e^{K(\Phi, \Phi^{\dagger})/M_P^2}\Big[ \big( D_{\Phi_i} W(\Phi) \big)
K^{\Phi_i \ov{\Phi}_j} \big(D_{\ov{\Phi}_j} W^*(\Phi^{\dagger}) \big)
 - \frac{3}{M_P^2} \left| W(\Phi) \right|^2 \Big],
\end{equation}
where $\Phi$ denotes the scalar field in general, the subscript means the derivative with respect to the field, $F_\Phi \equiv D_{\Phi} W = W_{\Phi} + K_{\Phi} W/M_P^2$, and 
$K^{\Phi_i \ov{\Phi}_j}$ is the inverse matrix of $K_{\Phi_i \ov{\Phi}_j}$. Here and hereafter, 
we neglect the contribution from the D-term. In our argument, we consider only the flat 
direction $\phi$ and the inflaton $I$ with $W=W(\phi)+W(I)$.

During inflation the scalar potential is dominated by the energy of inflaton. We can thus write 
the effective potential of the inflaton as
\begin{equation}
V(I) \simeq e^{K(I,I^{\dagger})/M_P^2}  \Big[ \big( D_I W(I) \big)
K^{I \bar{I}} \big(D_{\bar{I}} W^*(I^{\dagger}) \big)
 - \frac{3}{M_P^2} \left| W(I) \right|^2 \Big].
\end{equation}
In order to have positive potential energy, the first term in the parenthesis dominates:
$|D_I W(I)| \gtrsim |W(I)|/M_P$.
Since the total energy density is dominated by the inflaton, we can relate it to the Hubble 
parameter as $V(I) \simeq 3 H^2 M_P^2$. In the inflaton oscillation dominated era after inflation,
the same formula is applicable if one regards $I$ as its amplitude. For $|I| \sim M_P$, we have
$D_I W(I) \sim HM_P$ and $W(I) \lesssim  H M_P^2$.

The negative 
Hubble-induced mass terms should exist well after inflation until $H \sim m_\phi$, so we must 
thus seek for the case with $|I| \ll M_P$, even when $|I| \sim M_P$ during inflation. It is then 
necessary to equip nonminimal K\"ahler potential, because the minimal K\"ahler potential always 
results in a positive Hubble-induced mass term, which is shown shortly. In this case, 
we have $K^{I\bar{I}}\simeq 1$, $|F_I| \simeq HM_P$, and $|W(I)| \ll HM_P^2$.  

Now we consider if the Hubble-induced mass terms become positive or negative
for $|I| \ll M_P$. We take
the following five (the minimal and four nonminimal) K\"ahler potentials as typical examples:
\begin{eqnarray}
K_m & = & \phi^\dagger \phi + I^\dagger I, \\
\delta K_1 & = & \frac{a}{M_P^2} \phi^\dagger \phi  I^\dagger I, \\
\delta K_2 & = & \frac{b}{2M_P} I^\dagger \phi \phi + h.c., \\
\delta K_3 & = & \frac{c}{4M_P^2} I^\dagger I^\dagger \phi \phi + h.c., \\
\delta K_4 & = & \frac{d}{M_P} I \phi^\dagger \phi + h.c.
\end{eqnarray}

For the minimal K\"ahler potential, only cases (a) and (b) are nonzero. As is well known,
in this case, the Hubble-induced mass term has positive coefficient:
\begin{equation}
c_H = 3 + \left(\frac{e^{K(I,I^\dagger)} |F_I|^2}{V(I)} -1\right) \simeq 3,
\end{equation}
where the last equality holds for $|I| \ll M_P$.

Therefore, nonminimal K\"ahler potential should be sought for obtaining negative Hubble-induced
mass terms. In each case we consider, we obtain the Hubble-induced mass term 
$c_H H^2|\phi|^2$ with 
\begin{equation}
\label{Hmass}
c_H \simeq \left\{ \begin{array}{lcl} 
3(1-a) & & {\rm for} \ K_m + \delta K_1, \\[2mm] 
3(1+b^2) & &{\rm for} \ K_m + \delta K_2, \\[2mm]
3 & &{\rm for} \ K_m + \delta K_3, \\[2mm]
3(1+d^2) & & {\rm for} \ K_m + \delta K_4, \\
\end{array}  \right.
\end{equation}
for $|I| \ll M_P$.
The only possibility for a negative Hubble-induced mass term is introducing $\delta K_1$
with $a >1$.

If this is the only way for getting large VEVs during and after inflation, one may not seem
it very natural to have a successful Affleck-Dine mechanism. However, we show below that
large enough A-terms could lead the field to acquire large VEVs.

\section{Large VEV by Hubble-induced A-terms}
In this section, we describe how the effective potential acquires the minima at the large VEV
due to Hubble-induced A terms, even if the Hubble-induced mass term is positive. Considering 
only the second and third lines of Eq.(\ref{pot}), and rewriting as 
$\phi = \varphi \, e^{i\theta}/\sqrt{2}$, we have the potential of the form
\begin{equation}
V(\varphi) = \frac{1}{2} c_H H^2 \varphi^2 
+  \lambda^2 \frac{\varphi^{2(n-1)}}{2^{n-1}M_P^{2(n-3)}}
+ \ a_H H \frac{\varphi^n}{2^{\frac{n}{2}-1}nM_P^{n-3}}\cos(n\theta).
\end{equation}
For our purpose to obtain the minimum at large VEV, it is sufficient to set $\cos(n\theta)=-1$,
and consider only the particular radial direction with $n\theta=\pi$. It is then obvious that the 
$\varphi$ develops another minimum at $\varphi_{\rm min} \sim (H M_P^{n-3})^{1/(n-2)}$, 
provided that the following condition is met:
\begin{equation}
a_H^2 >  4(n-1) \lambda^2 c_H.
\end{equation}
Since the curvature at this minimum is of order $H^2$, the field rapidly settles down there 
during inflation. One might worry when this minimum is a local minimum. However, the transition
rate is approximately $P \sim \exp(-M_P^4/V(\phi)) \ll 1$ unless the dip and hill are extremely 
degenerate. Of course, one can set a little more severe condition 
$a_H^2 > n^2\lambda^2 c_H$, to make the dip as a global minimum. In any case, chaotic 
condition in the early inflationary stage will make the Affleck-Dine field settle into the minimum
at a large VEV with of order $O(1)$ probability.

The evolution of the field is very similar to that in the case of the negative Hubble-induced mass
term, since the field value of the newly developed minimum is almost the same if the parameters
such as $a_H$, $c_H$ and $\lambda$ are of order unity: 
$\varphi_{A, {\rm min}} \sim (H M_P^{n-3})^{1/(n-2)}$. After the field stuck into the 
minimum during inflation, it will stay there until $H\sim m_\phi$ when the Hubble-induced 
A term is overcome by the usual A-terms due to SUSY breaking by hidden sector. Thus, the 
field starts oscillation around the origin, and simultaneously feels torque to move along the 
phase direction. Since the field value and the power of the torque at the onset of the oscillation 
is the same as in the case of negative Hubble-induced mass term, the produced baryon number 
at that time is estimated as, for $O(1)$ difference in the phases,
\begin{equation}
n_B \sim \frac{A}{m_\phi} \frac{\varphi_{A,{\rm min}}^n}{M_P^{n-3}} 
\sim \left(\frac{m_\phi}{M_P}\right)^{\frac{n}{n-2}} M_P^3, 
\end{equation}
which is the same order of magnitude as Eq.(\ref{baryon}). Thus, the following evolution of the 
field should be similar, and hence we obtain almost the same amount of the baryon asymmetry 
of the universe.

For $|I| \ll M_P$, the minimal K\"ahler potential leads only to vanishing Hubble-induced A-terms,
so nonminimal ones are necessarily required, not only for developing minima at large VEV but
for obtaining the dynamical CP violation. The only nonvanishing Hubble-induced A terms appear
for $\delta K_2$ and $\delta K_4$ among which we considered:\footnote{
Here we correct these formulae in Ref.~\cite{KK}. $W(\phi)$ in Eq.(A12) in Ref.~\cite{KK} 
should be replaced by $W_\phi \phi$. } 
\begin{eqnarray}
- \sqrt{3} b \, W_\phi \, \phi^\dagger H + h.c. & \qquad & {\rm for} \ \delta K_2, \\
- \sqrt{3} d \, W_\phi \phi H + h.c. & \qquad & {\rm for} \ \delta K_4. 
\end{eqnarray}
Although the A terms of the case with $\delta K_2$ look a bit weird, the abilities to have minima
at large amplitude and CP violation is the same. The only difference is that minima in the phase 
direction are fewer by two. In either case, the minima will appear for 
$|b|, |d| > 2(n-1)^{1/2}/(n-2)$.

For more general case that $q\ne n$ in Eq.(\ref{pot}) with all the coefficients being $O(1)$, 
it is easy to see that $q < n$ is required in order to have (local) minima at large VEV. Let
the $\phi$-dependence as $K \propto \phi^r +h.c.$ and $W \propto \phi^n$. Since the A-term
proportional to $H$ comes from $\ds{(D_\phi W) K^{\phi \bar{I}} (D_{\bar{I}} W^*)+h.c.}$, it is
proportional to $\phi^{r+n-2}$. This implies $q=r+n-2$, leading to $r=1$. However, gauge-invariant
monomials in MSSM is constructed at least by two fields. Therefore, it works only for case with 
$q=n$ for sufficiently large coefficient in the A-term.

Since $q$ must not be larger than $n$, the K\"ahler potential should have only two
$\phi$s, namely, $\phi \phi$ or $\phi^\dagger \phi$. Also $K^{\phi \bar{I}}$ should not depend
upon the inflaton field $I$ in order to have non-vanishing A-term for $|I| \ll M_P$. Thus, 
$\delta K_2$ and $\delta K_4$ are the only working examples for obtaining the minima in the
large field value regime.

For completeness, let us consider the A-term proportional to $H^2$. As a result, it works only 
for such a term as $H^2 \phi^2 +h.c.$. This can be realized by 
$\ds{\delta K=\frac{g}{2M_P^2} I^\dagger I \phi \phi +h.c.}$ with $g>C_H \simeq 3$.

\section{Conclusions}
We have shown how the Affleck-Dine baryogenesis works in the context of supersymmetric
theory. Special attention is paid to the initial condition of the Affleck-Dine field which has to
have a large VEV during and after inflation. In the usual situations, the large VEV is achieved
by a negative Hubble-induced mass term due to SUSY breaking by the finite energy density
of the inflaton. We seek for the origin of the negative Hubble-induced mass terms for
various K\"ahler potentials. 

Most important fact that we have found here is that the minima at large VEV can be obtained by
large enough Hubble-induced A terms, even if the Hubble-induced mass term is positive. Since
A terms have minima irrespective of the signature of the coupling in the nonminimal K\"ahler
potential, it is robust for the Affleck-Dine field to have large VEV during and after inflation.
Thus, the Affleck-Dine mechanism for baryogenesis works in broader classes of theories.

\section*{Acknowledgments}
The author is grateful to M. Kawasaki for useful discussions. 
The work of S.K. is supported by the Grant-in-Aid for Scientific Research from the
Ministry of Education, Science, Sports, and Culture of Japan, No.~17740156.


\end{document}